\documentclass[journal=jpccck,manuscript=article]{achemso}
\usepackage[version=3]{mhchem}
\usepackage{graphicx}
\usepackage{amsmath}

\author{E. Durgun}
\affiliation{European Multifunctional Materials Institute (EMMI)}
\affiliation{Physique Theorique des Mat\'{e}riaux, Universit\'{e} de Li\`{e}ge (B5), B-4000 Li\`{e}ge, Belgium}
\affiliation{Department of Materials Science and Engineering, Massachusetts Institute of Technology, Cambridge, Massachusetts 02139, USA}
\email{edurgun@mit.edu}
\author{D. I. Bilc}
\affiliation{Physique Theorique des Mat\'{e}riaux, Universit\'{e} de Li\`{e}ge (B5), B-4000 Li\`{e}ge, Belgium}
\affiliation{Mol $\&$ Biomol Phys Dept, Natl Inst Res $\&$ Dev Isotop $\&$ Mol Technol, RO-400293 Cluj Napoca, Romania}
\author{S. Ciraci}
\affiliation{Department of Physics, Bilkent University, Ankara
06800, Turkey}
\affiliation{UNAM-Institute of Materials Science
and Nanotechnology, Bilkent University, Ankara 06800, Turkey}
\author{Ph. Ghosez}
\affiliation{European Multifunctional Materials Institute (EMMI)}
\affiliation{Physique Theorique des Mat\'{e}riaux, Universit\'{e} de Li\`{e}ge (B5), B-4000 Li\`{e}ge, Belgium}

\title[sinw]
{Interstitial Transition Metal Doping in Hydrogen Saturated Silicon Nanowires}

\begin{document}

\begin{abstract}
We report a first principles systematic study of atomic, electronic, and magnetic properties of hydrogen saturated silicon nanowires (H-SiNW) which are doped by transition metal (TM) atoms placed at various interstitial sites. Our results obtained within the conventional GGA+U approach have been confirmed using an hybrid functional. In order to reveal the surface effects we examined three different possible facets of H-SiNW along [001] direction with a diameter of $\sim$2nm. The energetics of doping and resulting electronic and magnetic properties are examined for all alternative configurations. We found that except Ti, the resulting systems have magnetic ground state with a varying magnetic moment. While H-SiNWs are initially non-magnetic semiconductor, they generally become ferromagnetic metal upon TM doping. Even they posses half-metallic behavior for specific cases.  Our results suggest that H-SiNWs can be functionalized by TM impurities which would lead to new electronic and spintronic devices at nanoscale.

\end{abstract}

\section{Introduction}
Among the research for building blocks in nano-devices, silicon nanowires (SiNW) are attracting increasing interest due to their remarkable physical, electronic, thermal, and chemical properties\cite{morales,holmes,boukai,cui}. Currently, rodlike, oxidation resistant SiNWs can be fabricated with a diameter down to 1 nm where quantum confinement effects are also observable\cite{ma1,*ma2,holmes}. The compatibility with current silicon based technology\cite{cui2,lauhon} makes SiNWs even more attractive and they can enter various device applications such as field effect transistors\cite{cui3}, light emitting diodes\cite{huang2}, lasers\cite{duan}, nanosensors\cite{hahm}, etc. The advances in synthesis will possibly increase the number of potential applications into various other fields.

At the theoretical level, bare and hydrogen saturated silicon nanowires (H-SiNW) have been extensively analyzed by using first principles calculations\cite{zhao,vo,rurali,rurali2,rurali3,engin1,engin2}. It is shown that while bare SiNWs are in general metallic, they become semiconductor (insulator) when saturated with hydrogen\cite{zhao,engin1,engin2}. The electronic band gap can be engineered by varying diameter, growth direction and cross section due to the quantum confinement effects\cite{ma1,*ma2,zhao}. More recently, superlattice structures leading to confined states induced by diameter modulated SiNWs\cite{yumur} and merged Si and Ge nanowires\cite{engin3} have been predicted.

Furthermore, growing research interest has also been devoted to functionalization of SiNWs with various dopants in order to study the chemical and biological sensitivities\cite{cui4,zhou,hahm}. The n- and p-doped H-SiNW  can provide excess carriers required in device applications like diodes and transistors\cite{singh,fernandez,engin4}. Recently, Wu et al.\cite{wu} have reported the room temperature magnetism of Mn-implanted SiNWs which opens the field to spintronic applications through transition metal (TM) doping. Theoretically Mn impurities in [111] SiNWs\cite{giorgi} and also external adsorption of various TM atoms on H-SiNWs are reported\cite{engin1,engin2}.

In this paper we present an extensive first principle study on hydrogen saturated silicon nanowires which are doped by transition metal atoms (Ti, V, Cr, Mn, Fe, and Co) placed at various interstitial sites. The energetics of doping and resulting electronic and magnetic properties are examined for all alternative configurations. In order to reveal surface effects, three different facets along [001] direction are considered. The magnetic ground state is determined by considering non-magnetic, ferromagnetic, and anti-ferromagnetic configurations. We also considered calculations by using a new type of hybrid functionals which in turn are compared with the results obtained by GGA+U.

\section{Methodology}

We have performed first-principles plane wave calculations within density functional theory(DFT)\cite{kohn1,*kohn2} implemented in {\sc vasp} code \cite{vasp1,*vasp2}. All calculations for non-magnetic, ferromagnetic and anti-ferromagnetic states are carried out by using the projector augmented wave (PAW) potentials.\cite{paw1,paw2} The exchange correlation potential has been approximated by generalized gradient approximation (GGA)\cite{gga}.

To model our silicon nanowires, we have used a supercell approach, including a 10 \AA\ vacuum space between wire replica along the two directions perpendicular to the wire axis. For doped nanowires, the impurity TM atom is periodically repeated along the [001] direction corresponding to the wire axis. In most calculations, one TM atom is introduced per basic wire unit, corresponding to a distance between consecutive impurities of 5.4 \AA.  Even we are not aiming to obtain impurity levels, in order to test the size dependence of our results, for specific cases we doubled our supercell size along [001] direction, so doubling the distance between an impurity TM atoms. The results were reasonably similar,\cite{Convergence}  confirming the main trends and our conclusions. As expected, E$_\textrm{b}$ slightly increases with lengthening the supercell size owing to the reduction in TM-TM interaction (which enhances TM-SiNW interaction), but electronic band structure profiles are not affected.

In the self-consistent potential and total energy calculations the Brillouin zone of supercell is sampled in the \textbf{k}-space within Monkhorst-Pack scheme\cite{monk} by (1x1x15) mesh points. A plane-wave basis set with a kinetic energy of 500 eV has been used. All atomic positions and lattice constant along the wire axis are optimized by using the conjugate gradient method where total energy and atomic forces are minimized. The convergence in energy is chosen as 10$^{-5}$ eV between two ionic steps, and the maximum force allowed on each atom is 0.01 eV/\AA.

A second set of calculations has been performed using the so-called B1-WC hybrid functional \cite{daniel}, which mixes 16\% of Hartree-Fock exchange with Wu-Cohen GGA \cite{wc} within B1 scheme \cite{b1}. These calculations have been performed using the linear combination of atomic orbitals method, as implemented in the {\sc crystal} code \cite{crystal}. The B1-WC hybrid functional used in this work was developed earlier \cite{daniel} and it was tested for H-SiNWs \cite{sinw-b1wc}. We used localized Gaussian-type basis sets including polarization orbitals and considered all the electrons for Si \cite{Porter}, H \cite{Gatti}, Cr \cite{Catti} and Fe \cite{Moreira}. The Brillouin zone integrations were performed using a 1x1x15 mesh of k points and a secondary 2x2x30 Gilat k-mesh was used in the calculation of Fermi energy and density matrix.  The self-consistent-field calculations were converged until the energy changes between interactions were smaller than 10$^{-8}$ Hartree using a Fermi smearing of 0.00032 Hartree ($\sim$100 K). An extra-large predefined pruned grid consisting of 75 radial points and 974 angular points was used for the numerical integration of charge density. The level of accuracy in evaluating the Coulomb and exchange series is controlled by five parameters \cite{crystal}: the values used in our calculations are 7, 7, 7, 7, and 14. The B1-WC calculations have been performed for the optimized structures obtained with the {\sc vasp} code.

\begin{figure}
\includegraphics[scale=0.6]{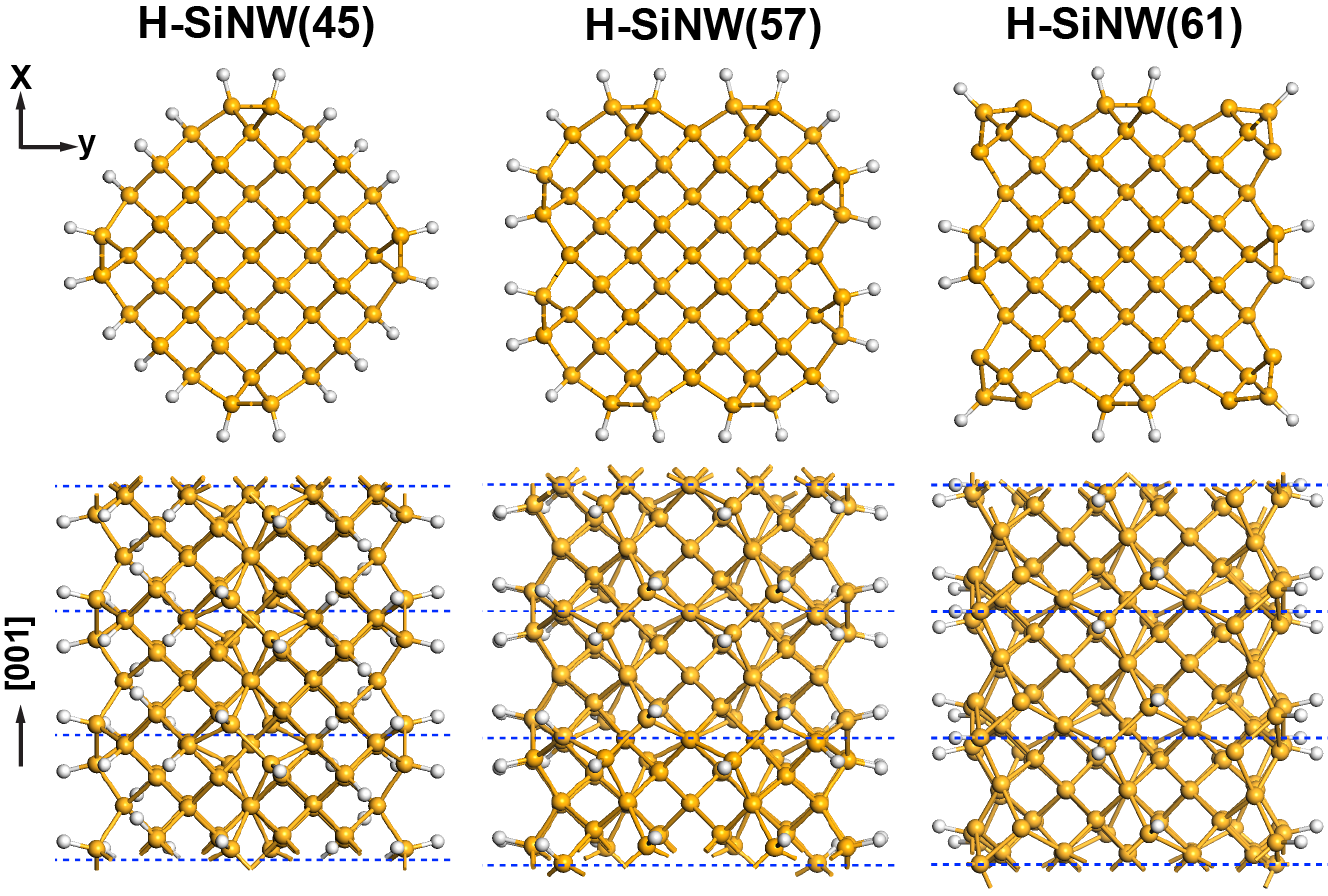}
\caption{(Color online) The top and side view of hydrogen saturated silicon nanowires along [001] direction with different facets for N=45,57 and 61 where N corresponds to number of silicon atoms in the unitcell of nanowire.} \label{fig:hsinw}
\end{figure}

\section{Hydrogen Saturated  Silicon Nanowires}

\begin{figure}
\includegraphics[scale=0.55]{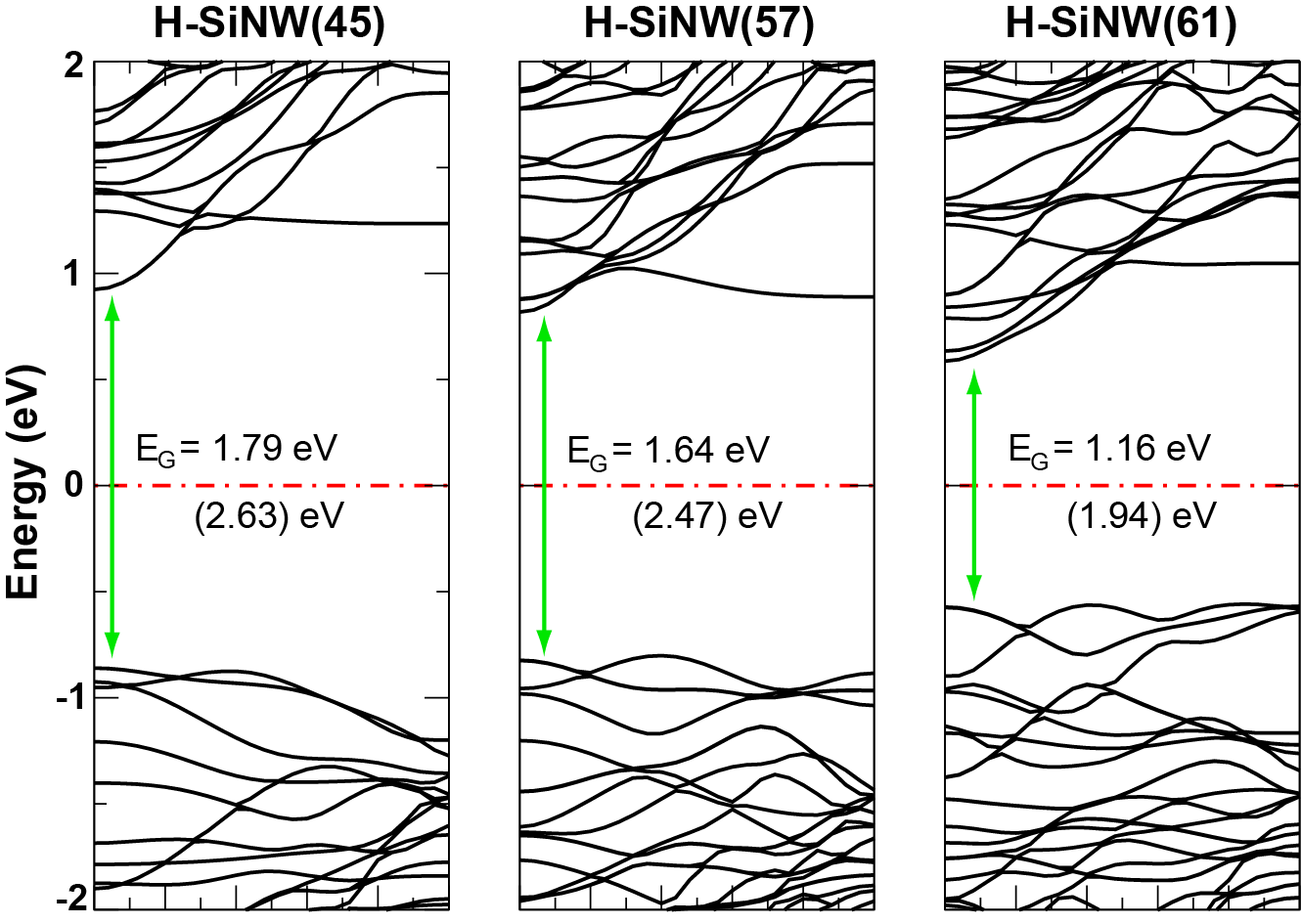}
\caption{(Color online) The band structures of hydrogen saturated silicon nanowires along [001] direction for N=45,57, and 61. The electronic energy band structures and the gaps (E$_\textrm{G}$) are calculated using GGA. The values given in parenthesis are obtained using B1-WC calculations.} 
\label{fig:hsinw_band}
\end{figure}

Before considering interstitial TM doping, we studied structural and electronic properties of prototype bare [SiNW(N)] and hydrogen saturated [H-SiNW(N)] silicon nanowires for N=45, 57, 61 where N is the number of silicon atoms in the unit cell of nanowires (\ref{fig:hsinw}). These H-SiNWs correspond to three possible facets along [001] direction and their diameter changes between 1.5-2.1 nm\cite{cao}. As already discussed in the introduction part, H-SiNWs as small as 1 nm have been fabricated\cite{ma1,*ma2} and at such small sizes edge effects become more important\cite{beigi,cao}.

The initial atomic positions of SiNWs are taken from silicon bulk crystal. Depending on the cross-section three alternative facets are possible along [001] direction. Upon ionic relaxation by minimizing both the total energy and forces on the atoms, the ground state configurations are obtained. The cohesive energy (E$_\textrm{C}$) of the SiNWs are calculated by the definition E$_\textrm{C}$=E$_\textrm{T}$[Si]-E$_\textrm{T}$[SiNW(N)/N] where E$_\textrm{T}$[Si] and E$_\textrm{T}$[SiNW(N)] is the total energy of free silicon atom and SiNW(N), respectively. E$_\textrm{C}$ is calculated as 4.91, 4.97, and 4.99 eV for N=45, 57, and 61, respectively.  Interestingly, SiNW(61) which has sharp corners  has the highest E$_\textrm{C}$ which is in agreement  with the results of Cao and coworkers\cite{cao}. Clearly, the presence of edges has a profound effect on the surface reconstruction of SiNW and thereby its electronic structure and stability\cite{beigi,cao}. Accordingly, while thick SiNWs prefers cylindrical or prism shape with a core that preserves diamond structure, the cross section with sharp corners becomes more favorable for very thin nanowires \cite{cao}.

As a next step, we saturate the dangling bonds of silicon atoms on the surface with hydrogen\cite{ma1,*ma2,engin1}. It is observed that while SiNWs are all metallic due to the surface states, they become semiconductors upon hydrogen saturation as shown in \ref{fig:hsinw_band}. The bang gap (E$_\textrm{G}$) tends to decrease from N=45 (E$_\textrm{G}$=1.79 eV) to N=61 (E$_\textrm{G}$=1.16 eV) due to an increase in diameter (despite the cross sections are different). This is consistent with predicted quantum confinement effects.

It is well known that while calculations based on DFT typically yield very accurate structural properties for silicon and other semiconductors, the standard approximations (LDA or GGA)  lead to a significant underestimation of the electronic band gaps due to intrinsic failure in handling self-interactions\cite{martin}. Accordingly we recalculated the band gaps by using B1-WC hybrid functional\cite{daniel}, the details of which are given in the methodology part. We noticed that the band structure profile that are obtained by GGA and B1-WC are very similar, and the correction results in a rigid shift of conduction and valence bands corresponding to a significant increase in the band gap. The corrected E$_\textrm{G}$'s are 2.53, 2.38 and 1.81 eV for N=45,57, and 61, respectively as shown in \ref{fig:hsinw_band}. The obtained values are in agreement with the experimentally available data\cite{ma1,*ma2} and theoretical GW corrected results \cite{zhao} for H-SiNWs within same diameter range\cite{sinw-b1wc}.

\section{Interstitial Transition Metal Doping}
\begin{figure}
\includegraphics[scale=0.5]{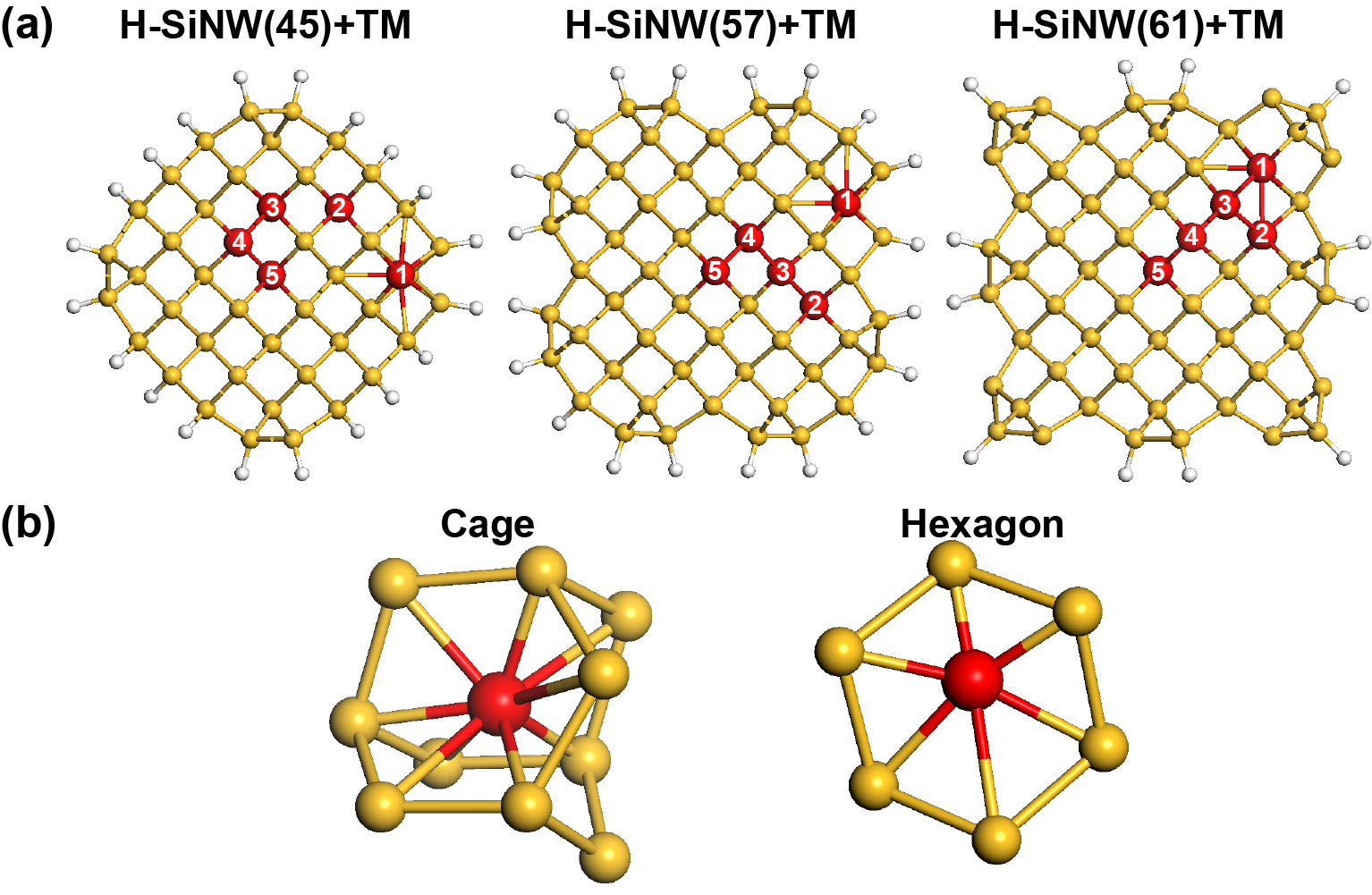}
\caption{(Color online) (a)The possible interstitial adsorption sites of transition metal (red spheres) atoms  labeled 1-5 in H-SiNW(N) for N=45,57 and 61 and (b) two alternative doping geometries.} \label{fig:tm_structure}
\end{figure}

Previously, we showed that TM atoms can strongly bind to H-SiNW(N) surfaces without deforming the nanowire structure\cite{engin1,engin2}. As the doping occurs at high temperatures there is a strong possibility that TM atom can also diffuse to interior regions. Accordingly we extend our previous analysis and consider various possible interstitial sites for six TM atoms (Ti, V, Cr, Mn, Fe, and Co) and analyze energetics, structural, electronic, and magnetic properties of TM doped H-SiNWs, labeled as H-SiNW(N)+TM(s), where TM is the type of transition metal atom and s is the interstitial doping site as shown in \ref{fig:tm_structure}. The obtained results, namely bond distances (d$_\textrm{Si-TM}$), binding energies (E$_\textrm{b}$), magnetic moments($\mu$) and the differences ($\Delta$E) between the total energies of spin-unpolarized (su) and spin-polarized (sp) states are summarized in \ref{tab:surface} and \ref{tab:core}.

\subsection{Structure and Energetics}
\begin{table*}
\caption{The bond distances, binding energy, magnetic moment per cell, total energy difference of polarized and unpolarized states for subsurface (s=1 and 2) sites. }
\label{tab:surface}
\centering
\resizebox{16 cm}{!}{%
\begin{tabular}{c|c|c|c|c|c|c|c|c|c|c|c|c|c}
&~s~~&\multicolumn{4}{c|}{N=45}&\multicolumn{4}{c|}{N=57}&\multicolumn{4}{c}{N=61}\\\hline
&&d(\AA)&E$_\textrm{b}$(eV)&$\mu$($\mu_b$)&$\Delta$E&d(\AA)&E$_\textrm{b}$(eV)&$\mu$($\mu_b$)&$\Delta$E(eV)&d(\AA)&E$_\textrm{b}$(eV)&$\mu$($\mu_b$)&$\Delta$E\\\hline
Ti&	1&	2.38-2.85&	3.42&	0.00&	0.00&	2.38-2.86&	3.59&	0.00&	0.00&	2.47-2.98&	4.34&	0.00&	 0.00\\
&	2&	2.43-2.84&	3.59&	0.00&	0.00&	2.45-2.80&	3.92&	0.00&	0.00&	2.42-2.85&	4.35&	0.00&	 0.00\\

V&	1&	2.34-2.77&	2.96&	1.01&	0.02&	2.34-2.78&	3.07&	0.83&	0.02&	2.45-2.83&	3.82&	2.97&	 0.02\\
&	2&	2.42-2.78&	3.09&	1.21&	0.02&	2.39-2.75&	3.38&	1.01&	0.01&	2.39-2.78&	3.39&	1.39&	 0.02\\

Cr&	1&	2.29-2.73&	1.42&	3.75&	0.09&	2.31-2.97&	1.72&	3.17&	0.21&	2.42-2.80&	2.46&	2.94&	 0.46\\
&	2&	2.45-2.75&	1.54&	3.72&	0.20&	2.43-2.73&	1.87&	3.51&	0.26&	2.41-2.81&	1.89&	3.46&	 0.27\\

Mn&1&	2.27-2.77&	1.75&	2.53&	0.05&	2.27-2.86&	1.89&	2.44&	0.10&	2.40-2.77&	2.66&	2.39&	 0.42\\
&	2&	2.42-2.77&	1.93&	3.00&	0.35&	2.41-2.75&	2.21&	3.00&	0.42&	2.39-2.81&	2.22&	3.00&	 0.39\\

Fe&1&	2.23-2.74&	3.18&	1.55&	0.19&	2.22-2.85&	3.28&	1.46&	0.19&	2.35-2.74&	3.83&	1.39&	 0.35\\
&	2&	2.39-2.73&	3.26&	2.00&	0.35&	2.37-2.73&	3.45&	2.00&	0.31&	2.36-2.81&	3.44&	2.00&	 0.28\\

Co&1&	2.17-3.01&	3.94&	0.00	&	0.00&	2.19-2.97&	4.16&	0.00&	0.00&	2.30-2.96&	4.76&	 0.00&	0.00\\
&	2&	2.39-2.70&	3.80&	1.00&	0.08&	2.37-2.77&	3.99&	1.00&	0.03&	2.37-2.89&	4.02&	1.00&	 0.02\\
\end{tabular}}
\end{table*}

\begin{table*}
\caption{The bond distances, binding energy, magnetic moment per cell, total energy difference of polarized and unpolarized states for core (s=4 and 5) sites. Results indicated by '*' correspond to hexagonal symmetry (See the text).}\label{tab:core}
\centering
\resizebox{16 cm}{!}{%
\begin{tabular}{c|c|c|c|c|c|c|c|c|c|c|c|c|c}
&~s~~&\multicolumn{4}{c|}{N=45}&\multicolumn{4}{c|}{N=57}&\multicolumn{4}{c}{N=61}\\\hline
&&d(\AA)&E$_\textrm{b}$(eV)&$\mu$($\mu_b$)&$\Delta$E&d(\AA)&E$_\textrm{b}$(eV)&$\mu$($\mu_b$)&$\Delta$E(eV)&d(\AA)&E$_\textrm{b}$(eV)&$\mu$($\mu_b$)&$\Delta$E\\\hline
Ti&	3&	2.45-2.78&	3.65&	0.00&	0.00&	2.44-2.84&	3.86&	0.00&	0.00&	2.42-2.85&	3.95&	0.00&	 0.00\\
&	4&	2.45-2.78&	3.65&	0.00&	0.00	&	2.47-2.80&	3.75	&	0.00	&	0.00	&	2.45-2.79&	 3.72	&	0.00	&	0.00\\
&	5&	2.45-2.83&	3.72&	0.00	&	0.00	&	2.48-2.73&	3.57	&	0.00	&	0.00	&	 2.47-2.79&	3.78&	0.00	&	0.00\\

V&	3&	2.44-2.73&	3.20	&	2.52&	0.08&	2.42-2.80&	3.33&	1.36&	0.03&	2.39-2.78&	3.39&	 1.40&	0.02\\
&	4&	2.44-2.73&	3.20&	2.52&	0.08&	2.43-2.76&	3.24	&	1.67	&	0.03	&	2.42-2.77&	3.28	 &	1.32	&	0.09\\
&	5&	2.41-2.79&	3.19&	1.41&	0.02&	2.46-2.79&	3.19	&	2.42	&	0.12	&	2.40-2.80&	3.21	 &	2.05	&	0.03\\

Cr&	3&	2.45-2.75&	1.78&	4.00&	0.40&	2.42-2.79&	1.78&	3.61&	0.25&	2.41-2.81&	1.89	&	 3.46&	0.27\\
&	4&	2.45-2.75&	1.78&	4.00&	0.40&	2.45-2.76&	1.79	&	3.82&	0.34	&	*2.44-2.80&	*1.83&	 *4.00&	*0.34\\
&	5&	2.43-2.78&	1.73&	4.00&	0.33&	2.47-2.70&	1.77	&	4.00	&	0.38	&	2.32-2.46&	1.91	 &	4.00	&	0.38\\

Mn&3&	*2.37-2.39&	*1.45&	*3.00&	*0.62&	2.40-2.79&	2.18&	3.00&	0.47&	2.39-2.81&	2.22&	3.00&	 0.39\\
     &4&	*2.32-2.38&	*1.54&	*3.00&	*0.60&	2.42-2.77&	2.14&	3.00&	0.45&	*2.33-2.37&	*1.49&	 *3.00&	*0.35\\
     &5&	2.40-2.79&	2.09&	3.00&	0.52&	2.44-2.69&	2.06	&	3.00	&	0.36	&	*2.31-2.38&	 *1.61&	*3.00&	*0.32\\

Fe&3&  *2.26-2.38&	*2.74&	*2.00&	*0.44&	2.38-2.77&	3.41&	2.00&	0.38&	2.36-2.81&	3.44&	2.00&	 0.28\\
&	4&	*2.26-2.39&	*2.84&	*2.00&	*0.40&	2.39-2.76&	3.38	&	2.00	&	0.34	&	*2.27-2.37&	 *2.81&	*2.00&	*0.41\\
&	5&	2.38-2.78&	3.34&	2.00&	0.45&	2.42-2.69&	3.35	&	2.00	&	0.27	&	*2.26-2.39&	 *2.89&	*2.00&	*0.29\\

Co&3&	*2.28-2.34&	*3.60&	*1.00&	*0.06&	2.36-2.79&	3.92&	1.00&	0.07&	2.37-2.89&	4.02&	1.00&	 0.03\\
&	4&	*2.28-2.33&	*3.68&	*1.00&	*0.02&	2.37-2.82&	3.94	&	1.00	&	0.06	&	2.37-2.89&	3.64	 &	1.00	&	0.04\\
&	5&	2.37-2.75&	3.85&	1.00&	0.08&	2.42-2.70&	3.95	&	1.00	&	0.05	&	*2.28-2.34&	 *3.83&	*1.00&	*0.03\\
\end{tabular}}
\end{table*}

As a first step, we consider different positions from subsurface to core regions at which TM atoms can settle. The projection of these sites are shown in \ref{fig:tm_structure}. This corresponds to one impurity atom for 6 layers of H-SiNW (\ref{fig:hsinw}). For each site there are two possible geometries for TM doping: the center of a cage where TM atom is surrounded by ten silicon atoms and the center of hexagon where TM atom is in the middle of six silicon atoms forming an almost planar geometry as shown in \ref{fig:tm_structure}. For all the cases the cage geometry yields stronger E$_\textrm{b}$ due to an increase in coordination number of TM atom. For most of the cases the hexagon geometry is not stable and TM atom moves to center of silicon cage.  Accordingly we labelled the results corresponding to hexagon geometry with '*' and otherwise mentioned the discussions are for TM atoms that are bound in cage geometry.

TM atoms can settle inside H-SiNW(N) without deforming the wire structure for core sites (s=3-5) and no structural difference is observed for different facets (\ref{fig:tm_structure}). On the other hand, edge effects become more important for subsurface sites. While no significant structural modification is observed for H-SiNW(61)+TM system, the Si-Si bond is broken upon TM impurity for H-SiNW(45)+TM(1) and H-SiNW(57)+TM(1) as shown in \ref{fig:tm_structure}. The minimum and maximum d$_\textrm{Si-TM}$ ranges between 2.2-2.9 \AA (generally 2.4-2.8 \AA) and are summarized in \ref{tab:surface} and \ref{tab:core}, for subsurface and core sites, respectively.

The comparison of E$_\textrm{T}$ indicates that the bond-breaking at site(1) for H-SiNW(45) and H-SiNW(57) results in an increase in E$_\textrm{T}$ and makes it less favorable to other possible subsurface site (s=2) where there is no deformation. On the other hand for the case of H-SiNW(61), site(1) which is inside a sharp corner (edge), is the lowest energy configuration. For core sites, E$_\textrm{T}$'s slightly vary while moving from subsurface to the center (3$\rightarrow$5) and the influence of surface becomes less significant.

\begin{figure*}
\includegraphics[scale=0.6]{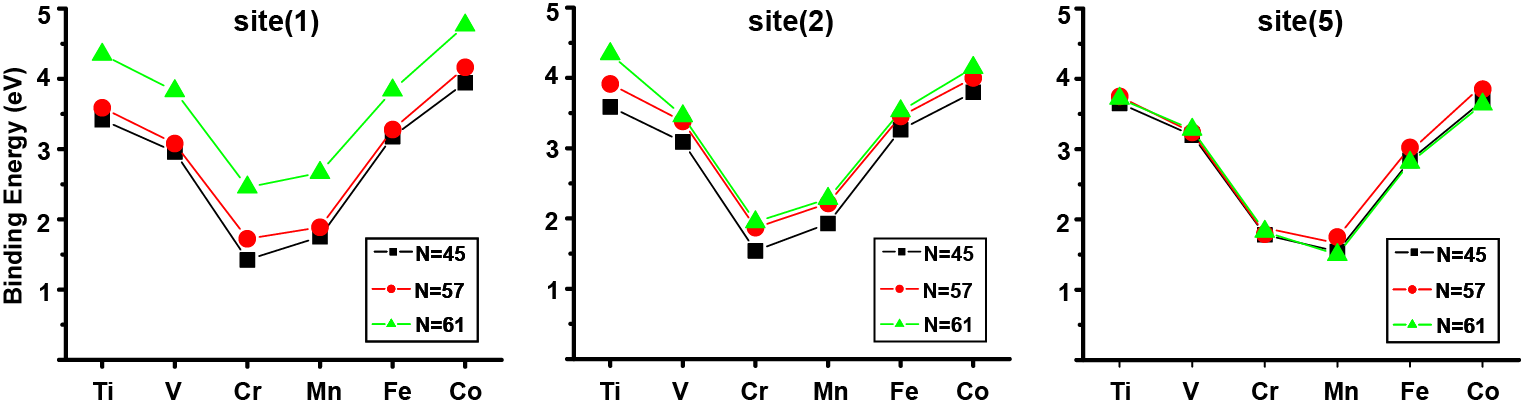}
\caption{(Color online) The binding energy trend with respect to type of transition metal atoms (Ti, V, Cr, Mn, Fe, and Co) in H-SiNW(N) for N=45, 57, and 61 for the sites 1, 2(subsurface), and 5(core).} \label{fig:binding}
\end{figure*}

We calculate the binding energy(E$_\textrm{b}$) of TM atoms for different sites by using the expression:

\begin{equation}
\textrm{E}_\textrm{b}= \textrm{E}_\textrm{T}\textrm{[H-SiNW(N)]+E[TM]-E}_\textrm{T}\textrm{[H-SiNW+TM]}  \label{func:binding}
\end{equation}
in terms of the total energy of optimized H-SiNW(N) and H-SiNW(N)+TM and the total energy of the string of TM atoms having the same lattice parameter as H-SiNW(N)+TM, all calculated in the same supercell. Interestingly, the variation of E$_\textrm{b}$ with type of TM atoms follows Friedel model\cite{friedel} for both different sites and facets (\ref{fig:binding})\cite{engin2003,engin2004}. Accordingly, the lowest E$_\textrm{b}$ is obtained for Cr and Mn and the highest E$_\textrm{b}$ is obtained for Co and Ti depending on the number of filled d-states. When different facets are compared, E$_\textrm{b}$ for N=61 is higher than N=57 and N=45 for subsurfaces sites, and there is no significant variation noticed for core sites. When doping sites are examined, interestingly the highest E$_\textrm{b}$ is obtained at site(5), site(2), and site(1) for H-SiNW(45), H-SiNW(57), and H-SiNW(61), respectively. The results point out that the structural properties as well as the energetics depend on both the type of TM and also the cross section due to surface effects.

\subsection{Magnetic Properties}

The total energies(E$_\textrm{T}$) for the considered systems are obtained both from spin-polarized (sp) and -unpolarized (su) states and the energy difference ($\Delta\textrm{E}=\textrm{E}_\textrm{T}^\textrm{su}-\textrm{E}_\textrm{T}^\textrm{sp}$) is used to determine the lowest energy configuration. According to this definition sp is the lowest energy configuration when $\Delta\textrm{E} > 0$. As shown in \ref{tab:surface} and \ref{tab:core}, expect for Ti, the lowest energy configuration is obtained for sp state. Additionally we also compared the E$_\textrm{T}$'s in ferromagnetic  and anti-ferromagnetic states and confirmed that the ground state is ferromagnetic except Ti which yields paramagnetic ground state. The magnetic moments, $\mu$ vary depending on both the type of TM and also doping site as listed in \ref{tab:surface} and \ref{tab:core}.

Using the energy difference between ferromagnetic and anti-ferromagnetic states, the Curie temperature (T$_\textrm{C}$) can be roughly estimated in the mean field approximation. Accordingly we consider H-SiNW(45)+TM as prototype, and calculated T$_\textrm{C}$ as  $\sim$600K ,900K , and 400K, for Cr, Mn, and Fe respectively which indicate that ferromagnetic state is stable over room temperature. On the other hand T$_\textrm{C}$ for V and Co is around $\sim$50K which shows that they can only have ferromagnetic ground state at very low temperatures. Finally Ti (and also Ni) yields paramagnetic ground state even at 0K. When the trend of T$_\textrm{C}$ is compared with the unpaired $d$-electrons ($d^{un}$) of TM dopants, a strong relation can be noticed. When $d^{un}$=2 (Ti and Ni), T$_\textrm{C}$=0 (in other words paramagnetic), $d^{un}$=3 (V, Co) T$_\textrm{C}\simeq50$, $d^{un}$=4 (Fe) T$_\textrm{C}\simeq400$ and $d^{un}$=5 (Cr and Mn), T$_\textrm{C} > 600$. Accordingly we can conclude that the stability of ferromagnetic state is mainly determined by the number of unpaired $d$-states.

\subsection{Electronic Structure}
\begin{figure*}
\includegraphics[scale=0.45]{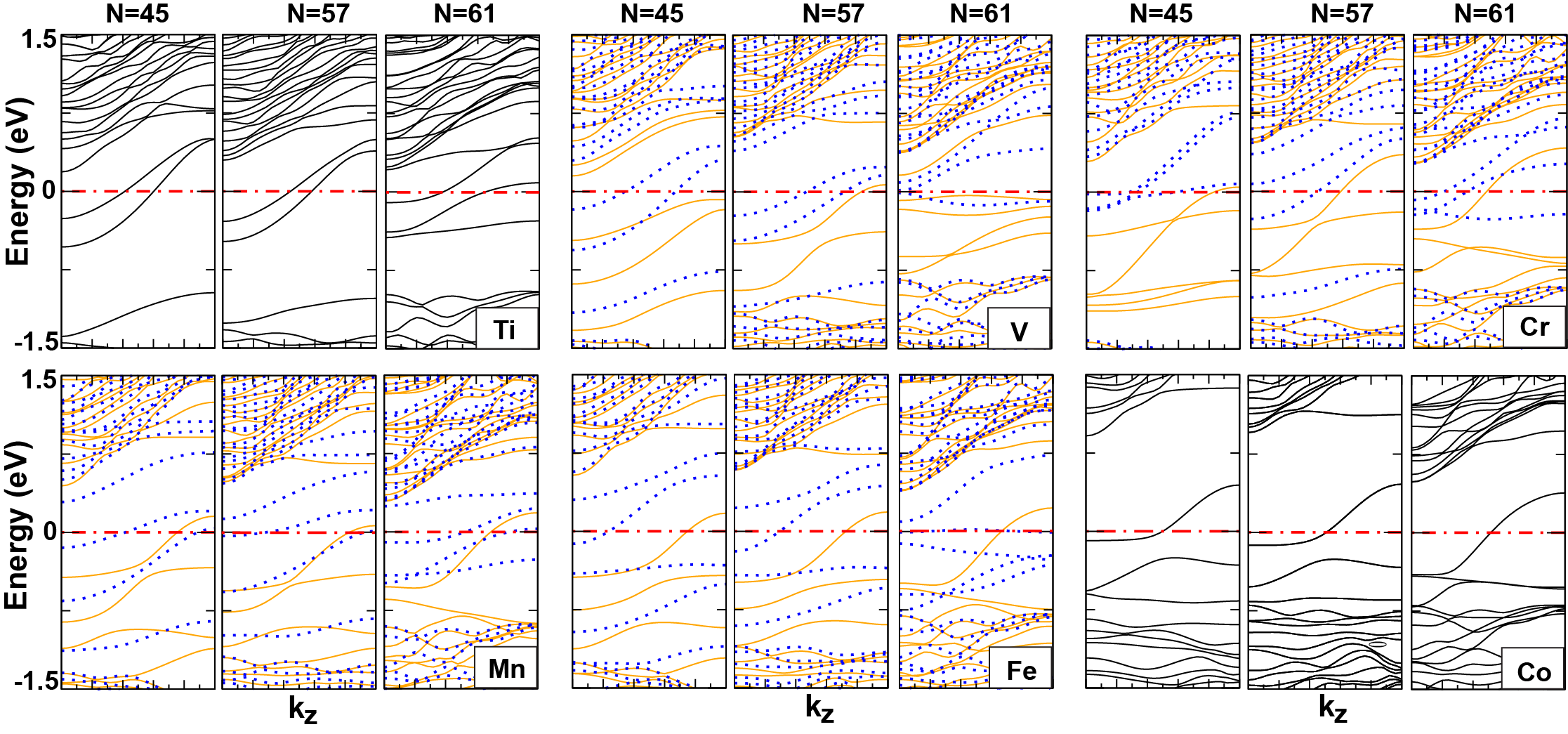}
\caption{(Color online) The electronic band structures of H-SiNW(N)+TM at site(1) for N=45,57, and 61 and TM=Ti, V, Cr, Mn, Fe, and Co. Solid and dotted lines indicate spin up and down states, respectively. } \label{fig:band_s1}
\end{figure*}

\begin{figure*}
\includegraphics[scale=0.45]{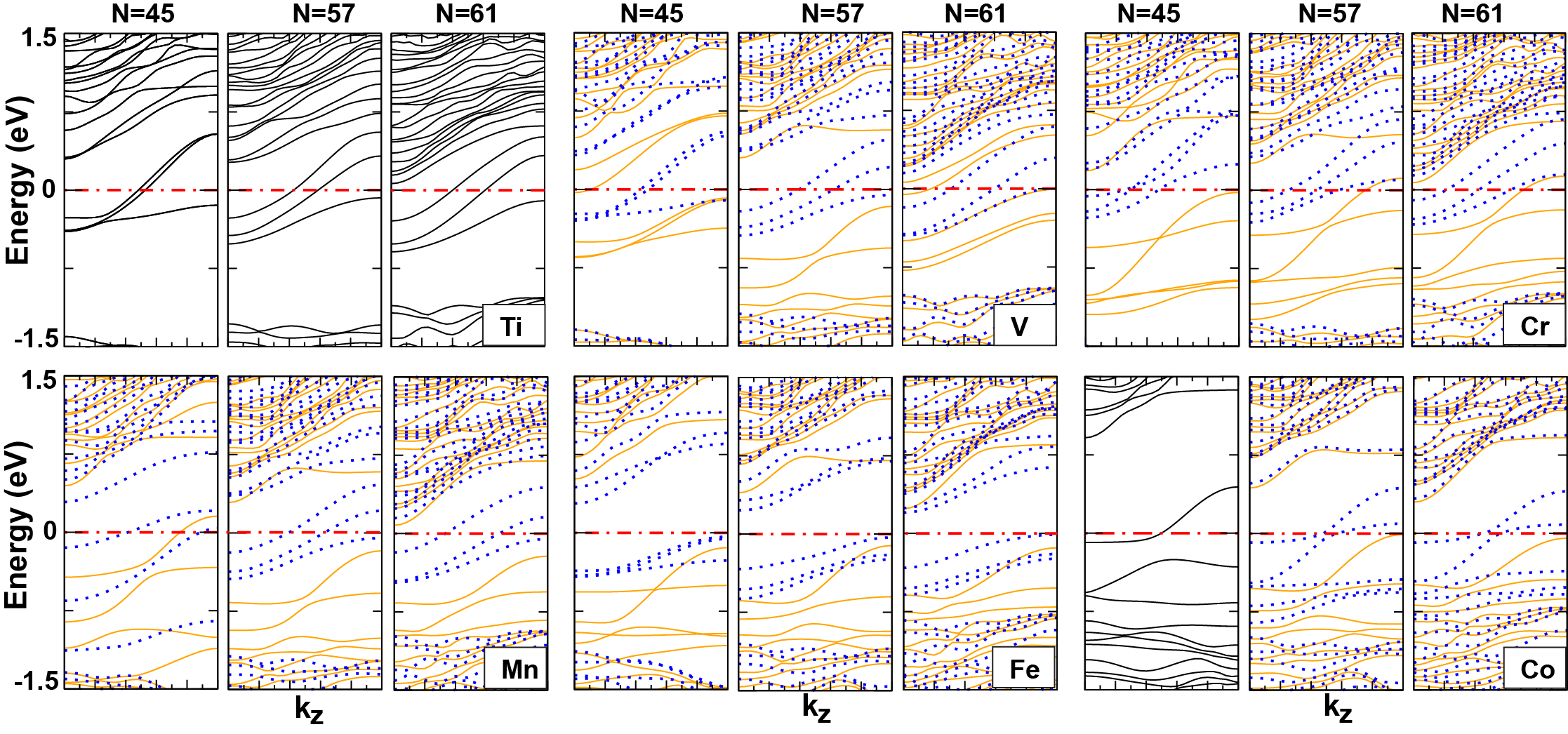}
\caption{(Color online) The electronic band structures of H-SiNW(N)+TM at site(2) for N=45,57, and 61 and TM=Ti, V, Cr, Mn, Fe, and Co. Solid and dotted lines indicate spin up and down states, respectively.} \label{fig:band_s2}
\end{figure*}

\begin{figure*}
\includegraphics[scale=0.45]{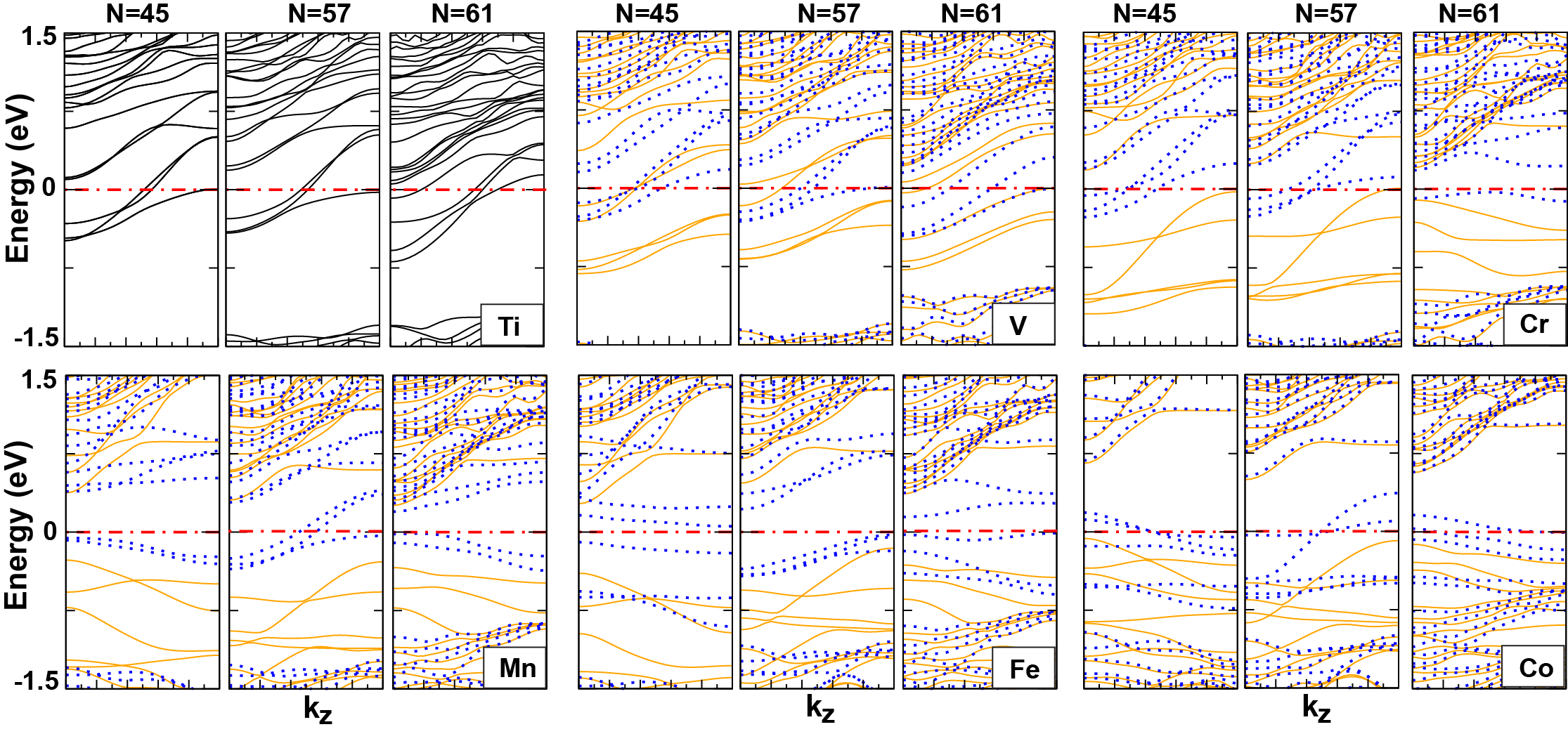}
\caption{(Color online) The electronic band structures of H-SiNW(N)+TM at site(4) for N=45,57, and 61 and TM=Ti, V, Cr, Mn, Fe, and Co. Solid and dotted lines indicate spin up and down states, respectively.} \label{fig:band_s4}
\end{figure*}

\begin{figure*}
\includegraphics[scale=0.45]{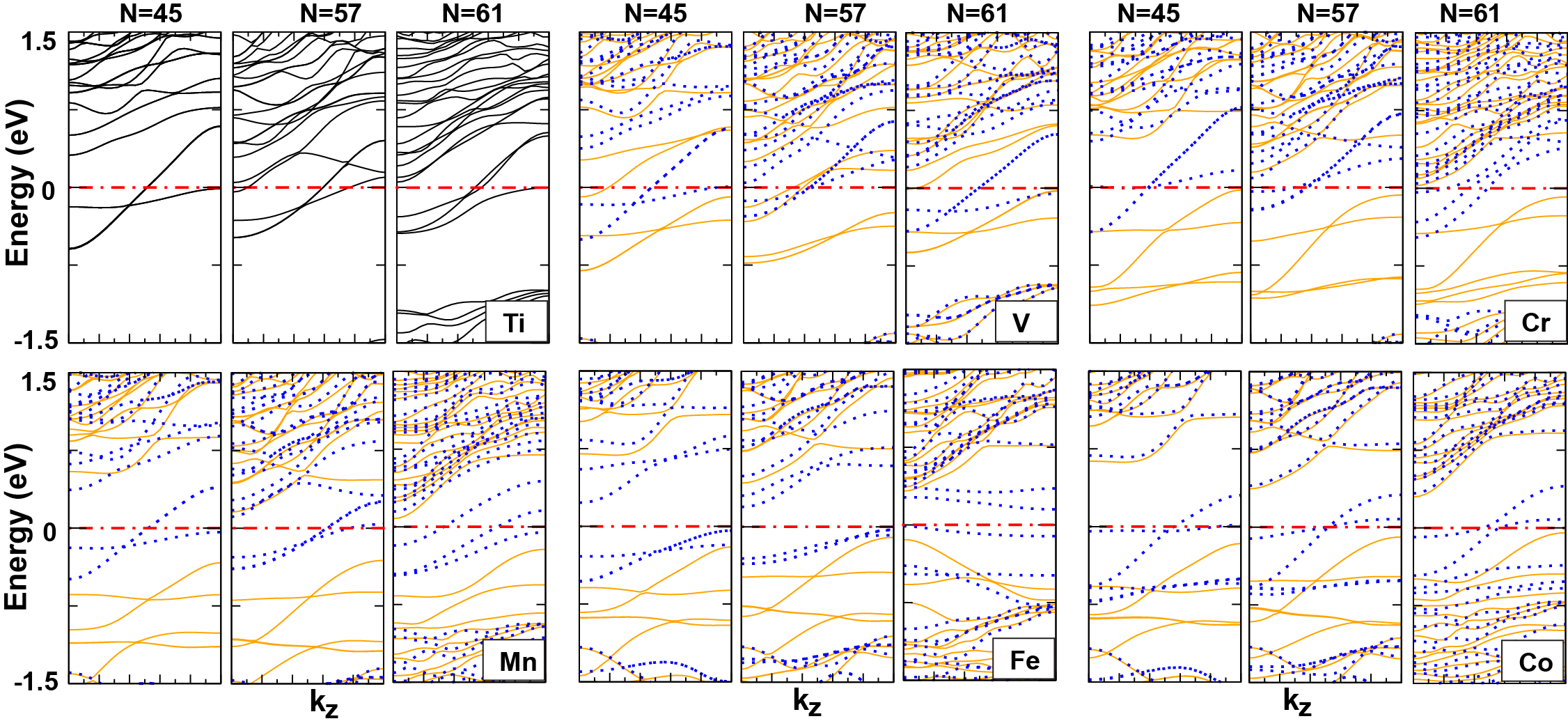}
\caption{(Color online) The electronic band structures of H-SiNW(N)+TM at site(5) for N=45,57, and 61 and TM=Ti, V, Cr, Mn, Fe, and Co. Solid and dotted lines indicate spin up and down states, respectively.} \label{fig:band_s5}
\end{figure*}

Our calculations reveal that electronic structures are strongly affected by the type of TM atoms, doping site and also cross section of H-SiNW(N). The resulting band structures for subsurface (s=1,2) and core sites (s=4,5) are shown in Fig\ref{fig:band_s1}-\ref{fig:band_s5}. Depending on the configuration the ground state can be paramagnetic or ferromagnetic and electronic structure can be metal, semiconductor, semi-metal or half-metal\cite{engin1,engin2}.

If we analyze the systems one by one, H-SiNW(N)+Ti configuration is paramagnetic metal for all considered systems. Interestingly, for V at site(1), the nanowires become half-metal, metal and semi-metal for N=45, 57, and 61, respectively and this indicates the effect of cross section on the electronic structure. Half-metallic state is also noticed for N=57 at site(2), and for all other doping sites (s=3-5), the system is ferromagnetic metal with varying magnetic moment. Cr and Mn behave very similarly. The nanowires are in general ferromagnetic metal for subsurface sites and they become half-metallic for core regions. Exceptionally H-SiNW(61)+Cr(4), H-SiNW(61)+Mn(4,5) are ferromagnetic semiconductor as TM is bound to hexagonal site instead of cage configuration (\ref{fig:tm_structure}). For Fe, all H-SiNWs are ferromagnetic semi-metal at site(1), but they become ferromagnetic semiconductor for all other sites including hexagonal doping site. And finally, for Co, all H-SiNWs are paramagnetic metal at site(1) and become half-metal or ferromagnetic semiconductor for core regions.

 Combining the results from the previous sections we can conclude that:

(i)The energetics of binding is almost identical at core sites for different types H-SiNWs but the variation is noticed for subsurface sites due to surface effects
(ii)The trend of E$_\textrm{b}$ follows Friedel model\cite{friedel} independent of H-SiNW type and binding site.
(iii)The electronic ground state  of H-SiNW(N)+TM is mainly determined by the type of (or the d-electron configuration of) TM dopants but can also differ for the same TM depending on the doping site and geometry.
 (iv)TM doping generally induces metallization (either paramagnetic or ferromagnetic) except for the case of Fe. H-SiNW(N)s still remain semiconducting after Fe doping with varying electronic band gap. The semi-metallic behavior of H-SiNW(N)+Fe(1) is an exceptional case and will be further discussed in the next section.
 (v) H-SiNW(N)+(Cr, Mn) systems are in general ferromagnetic metal for subsurface sites and they start to posses novel half-metallic behavior for core sites.
 (vi)However, H-SiNW(N)+V and H-SiNW(N)+Co also have half-metallic ground state for specific cases, they are ferromagnetic only at very low temperatures which make them practically paramagnetic metals. Accordingly, stable half-metallic ground state is unique to Cr and Mn where all the d-electrons are unpaired.
 vii)When Mn and Cr is settled  in the center of silicon hexagonal plane, d$_\textrm{Si-TM}$ becomes equal for all six nearest silicon atoms and it leads to different hybridization of p- (that belongs to H-SiNW) and d-orbitals (that belongs to TM) which makes the systems semiconductor instead of half-metal. However this configuration is energetically less favorable when compared to cage geometry but can be stable.

\section{Hybrid functionals and GGA+U}
It is well-known that the usual approximations to DFT, such as  LDA or GGA,  lead to significant underestimation of the band gaps due to intrinsic failure in handling self-interactions\cite{martin}. GW corrections\cite{hedin} appeared as a powerful tool  to overcome this problem and were recently applied to silicon nanowires\cite{zhao,bruno}. Unfortunately such GW calculations are computationally very costly and it is difficult to apply this technique on large systems. Moreover, GW corrections are performed on the frozen structure and do not allow for self-consistent structural relaxations.

\begin{figure}
\includegraphics[scale=0.6]{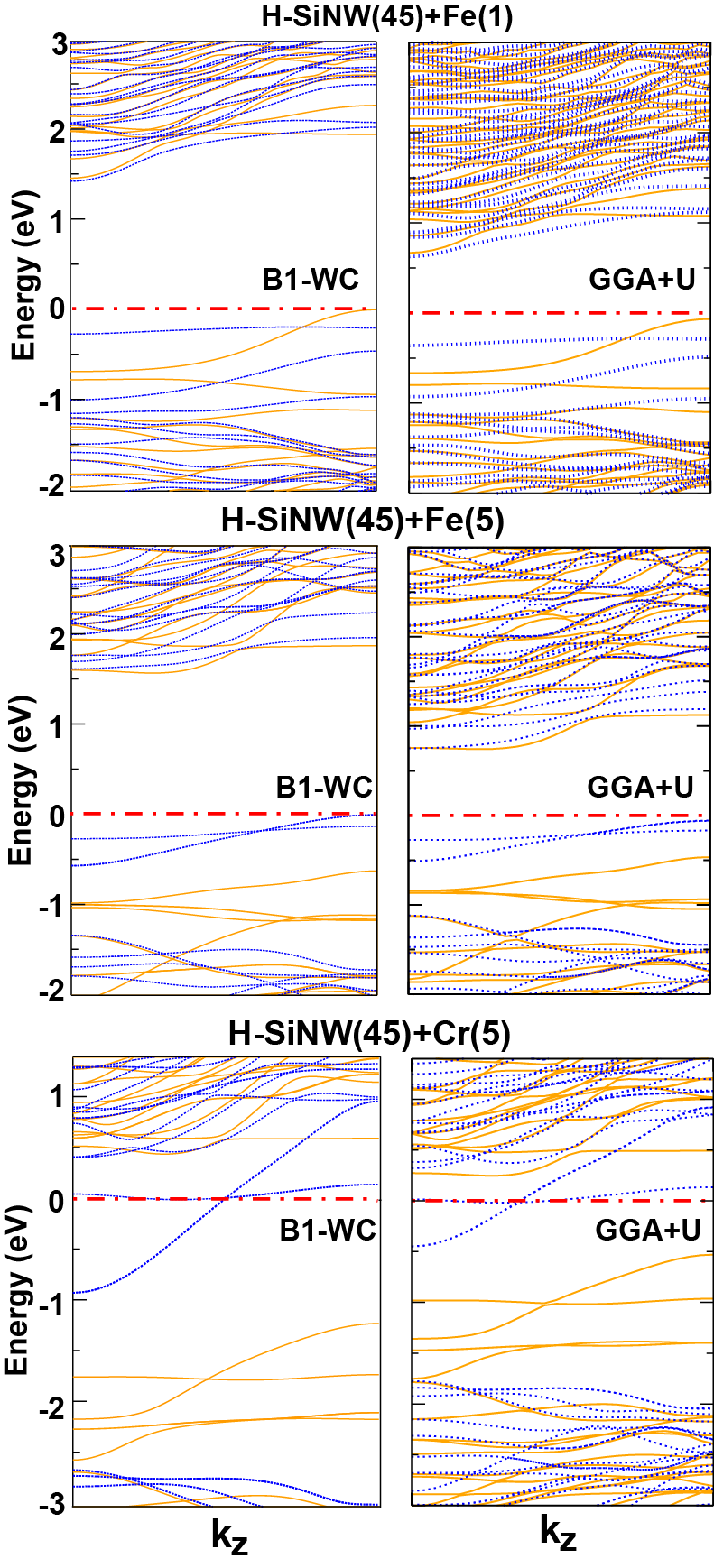}
\caption{(Color online) The electronic bands structures of H-SiNW(45)+Fe at site(1), H-SiNW(45)+Fe at site(5), and H-SiNW(45)+Cr at site(5) obtained by B1-WC and GGA+U calculations. Solid and dotted lines indicate spin up and down states, respectively.} \label{fig:hybrid}
\end{figure}

Alternatively, calculations using hybrid functionals\cite{becke1}, which basically combine Hartree-Fock and DFT, usually  provide improved electronic properties compared to DFT, although they are often less accurate for structural properties. Hybrid calculations are less costly than GW corrections and also allow for structural relaxations. Hybrid functionals were used to calculate the electronic structure of carbon nanotubes\cite{avramov,barone1,barone2,barone3} and graphene nanoribbons\cite{barone4},  and yield good agreement with experimental data. More recently, Rurali et al.\cite{rurali3} also applied hybrid functionals to silicon nanowires to accurately determine electronic band gap for varying diameter and orientation.

In this section, we compared the results obtained in the previous sections with those obtained by using B1-WC hybrid functional. We applied this technique only to specific cases, such as H-SiNW(45)+Fe at site(1) which is a ferromagnetic semimetal, H-SiNW(45)+Fe at site(5) which is a ferromagnetic semiconductor, and H-SiNW(45)+Cr [which is very similar to H-SiNW(45)+Mn] at site(5) which is a half-metal and compare the results obtained with GGA and GGA+U\cite{dudarev} calculations as shown in \ref{fig:hybrid}.

When B1-WC is used, the band structure of H-SiNW(45)+Fe at site(1) is altered and becomes semiconductor instead of semi-metal. Similar band profile with a smaller band gap is also obtained when Hubbard-U (U=3 eV) term is considered for the strong on-site 3d electronÐelectron interactions on Fe. Similar results are obtained for H-SiNW(45)+Fe at site(5). The system remains to be semiconducting but bang gap significantly increases when compared to GGA results. Accordingly we conclude that semi-metallic behavior of H-SiNW(N) is an artifact of DFT at GGA level and H-SiNW remains semiconducting  with a modified band gap upon Fe doping for both subsurface and core sites.

The novel half-metallic ground state of H-SiNW(45)+Cr at site(5) is preserved by B1-WC and also when U term is introduced. Interestingly, the dispersive metallic spin state ($\uparrow$) is not affected by either approaches but the the band gap for the insulating spin state ($\downarrow$) is modified. In other words, half-metallic ground states obtained at the DFT-GGA level\cite{engin1,engin2} are even more stable than predicted.

Finally, the results indicate that B1-WC is not only capable of correcting the under estimated electronic band gaps but also can handle strong on-site d electron-electron interactions without introducing Hubbard-U term, being a good alternative to LDA(GGA)+U for strongly correlated systems \cite{Marco,Alina}.

\section{Conclusions}

We analyzed the structural, electronic, and magnetic properties of hydrogen saturated silicon nanowires with interstitial transition metal doping. We found that the electronic and magnetic ground state is mainly determined by the type of (or d-electron configuration of) transition metal atom, it can also be  affected by doping site and cross section (surface effects) of silicon nanowire. Upon transition metal doping there is a tendency for metallization except for the case of Fe and even novel half-metallic configuration can be obtained for specific cases. The stability of ferromagnetic ground state and energetics of binding are inversely proportional and depends on the number of unpaired 3$d$-electrons following the Friedel model. The obtained results are also compared with those calculated using new type hybrid functional (B1-WC) in order clarify the limitations of density functional theory at GGA level and also to find out the capabilities of B1-WC. We believe that our detailed analysis will guide both experimental and theoretical studies related with doping of silicon nanowires and other systems and moreover present results hold the promise for the use of silicon nanowires in various spintronic applications upon transition metal doping.

\acknowledgement
This work was supported by the Interuniversity Attraction Poles Program (Grant No. P6/42)-Belgian State-Belgian Science Policy and the ARC project TheMoTher. Ph.G. acknowledges Research Professorship from the Francqui Foundation. D.B. thanks additional financial support from Romanian National Authority for Scientific Research, CNCS Ð UEFISCDI, project number PN-II-RU-TE-2011-3-0085.

\bibliography{references}

\begin{mcitethebibliography}{64}
\bibitem[Con()]{Convergence}
 For the H-SiNW(45) and impurity at site 5 (core), the computed binding
  energies for Ti, V, Cr, Mn, Fe and Co impurities are respectively of 3.72
  (3.83), 3.19 (3.32), 1.73 (1.83), 2.09 (2.17), 3.34 (3.70) and 3.85 (4.16) eV
  using a single (double) unit cell along the wire direction corresponding to
  distance between impurities of about 5.4 (10.8) \AA.\relax
\mciteBstWouldAddEndPunctfalse
\mciteSetBstMidEndSepPunct{\mcitedefaultmidpunct}
{}{\mcitedefaultseppunct}\relax
\EndOfBibitem
\end{mcitethebibliography}
\bibliographystyle{unsrt}

\end{document}